\documentclass[journal=nalefd,manuscript=letter]{achemso}
\setkeys{acs}{articletitle=false, maxauthors = 0}

\usepackage[T1]{fontenc} 
\usepackage{natbib}
\usepackage{textcomp}
\usepackage{graphicx} 
\usepackage{amssymb}
\usepackage{epsfig}
\usepackage{units}
\usepackage{caption}
\usepackage{lineno} 
\usepackage{hyperref}
\usepackage[noabbrev]{cleveref}
\usepackage{soul}
\usepackage{color}

\author{Adolfo De Sanctis}
\email{a.de-sanctis@exeter.ac.uk}
\affiliation[University of Exeter]{Centre for Graphene Science, College of Engineering, Mathematics and Physical Sciences, University of Exeter, Exeter EX4 4QF, United Kingdom}
\altaffiliation{Contributed equally to this work}
\author{Jake D. Mehew}
\affiliation[University of Exeter]{Centre for Graphene Science, College of Engineering, Mathematics and Physical Sciences, University of Exeter, Exeter EX4 4QF, United Kingdom}
\altaffiliation{Contributed equally to this work}
\author{Saad Alkhalifa}
\affiliation[University of Exeter]{Centre for Graphene Science, College of Engineering, Mathematics and Physical Sciences, University of Exeter, Exeter EX4 4QF, United Kingdom}
\alsoaffiliation[University of Duhok]{University of Duhok, Duhok, 42001 Kurdistan Region, Iraq}
\author{Freddie Withers}
\affiliation[University of Exeter]{Centre for Graphene Science, College of Engineering, Mathematics and Physical Sciences, University of Exeter, Exeter EX4 4QF, United Kingdom}
\author{Monica F. Craciun}
\affiliation[University of Exeter]{Centre for Graphene Science, College of Engineering, Mathematics and Physical Sciences, University of Exeter, Exeter EX4 4QF, United Kingdom}
\author{Saverio Russo}
\email{s.russo@exeter.ac.uk}
\affiliation[University of Exeter]{Centre for Graphene Science, College of Engineering, Mathematics and Physical Sciences, University of Exeter, Exeter EX4 4QF, United Kingdom}

\title[Strain in G/hBN heterostructures] {Strain-engineering of twist-angle in graphene/hBN superlattice devices}

\keywords{Graphene, hBN, superlattice, twist-angle, strain, Raman}

\begin{document}

\begin{tocentry}
	\includegraphics[]{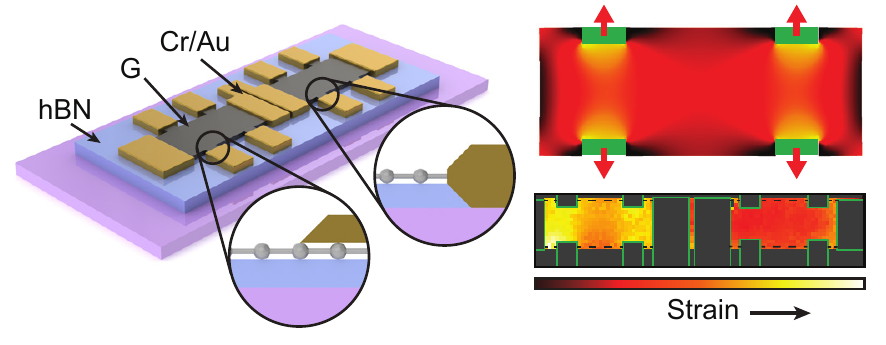}
\end{tocentry}

\begin{abstract}
	The observation of novel physical phenomena such as Hofstadter's butterfly, topological currents and unconventional superconductivity in graphene have been enabled by the replacement of SiO$_2$ with hexagonal Boron Nitride (hBN) as a substrate and by the ability to form superlattices in graphene/hBN heterostructures. These devices are commonly made by etching the graphene into a Hall-bar shape with metal contacts. The deposition of metal electrodes, the design and specific configuration of contacts can have profound effects on the electronic properties of the devices possibly even affecting the alignment of graphene/hBN superlattices.
	
	In this work we probe the strain configuration of graphene on hBN contacted with two types of metal contacts, two-dimensional (2D) top-contacts and one-dimensional (1D) edge-contacts. We show that top-contacts induce strain in the graphene layer along two opposing leads, leading to a complex strain pattern across the device channel. Edge-contacts, on the contrary, do not show such strain pattern. A finite-elements modelling simulation is used to confirm that the observed strain pattern is generated by the mechanical action of the metal contacts clamped to the graphene. Thermal annealing is shown to reduce the overall doping whilst increasing the overall strain, indicating and increased interaction between graphene and hBN. Surprisingly, we find that the two contacts configurations lead to different twist-angles in graphene/hBN superlattices, which converge to the same value after thermal annealing. This observation confirms the self-locking mechanism of graphene/hBN superlattices also in the presence of strain gradients. Our experiments may have profound implications in the development of future electronic devices based on heterostructures and provide a new mechanism to induce complex strain patterns in 2D materials.
\end{abstract}

\clearpage
The high charge carrier mobility attained at room temperature in graphene encapsulated in hexagonal Boron Nitride (hBN)\cite{Dean2010} has enabled the observation of ballistic transport over macroscopic distances\cite{Wang2013b,Banszeruse2015,Banszerus2016} holding the promise for the development of room-temperature electrical equivalents of optical circuits. As opposed to suspended graphene structures\cite{Khodkov2015} in which low-energy flexural phonons impose severe limitations on the maximum value of charge carrier mobility observable at room temperature\cite{Mariani2008,Castro2010}, in supported structures the optical phonons of the substrate play a central role. Compared to SiO$_2$, the optical phonons in hBN have higher energy and this results in an increase of the charge carrier mobility in graphene by an order of magnitude\cite{Dean2010}. Phonon scattering is not the only limiting factor to carrier mobility on SiO$_2$, scattering from adsorbates such as water and substrate roughness also dominate its value. Being free from dangling bonds and lattice matched to graphene within $\delta\sim1.7\%$, hBN also allows for an atomically clean interface to be formed. Crucially, the van der Waals attraction between these two-dimensional (2D) materials is strong enough to push contamination outside of the overlap region, resulting in an atomic-scale self cleaning mechanism which was shown to work for mechanically exfoliated flakes as well as large area graphene grown by chemical vapour deposition\cite{Kretinin2014,Banszeruse2015,Banszerus2016}. Another major breakthrough made possible by the encapsulation of graphene in hBN has been the realization of high-quality one-dimensional (1D) electrical contacts to graphene. In these edge-contact geometries, low temperature ballistic transport was reported over $15\,\mathrm{\mu m}$ together with substrate-phonon limited room-temperature charge carrier mobility\cite{Wang2013b}.

Moir\'e interference patterns are observed for graphene on hBN owing to the small lattice mismatch between the two crystals. The rotation of graphene with respect to the underlying hBN produces patterns each with a different Moir\'e wavelength,\cite{Xue2011,Decker2011} suggesting that effective periodic potentials are formed. For massless Dirac fermions this results in the formation of new Dirac points in the electronic band structure whose energy is determined by the Moir\'e wavelength.\cite{Yankowitz2012} Superlattice structures have led to the observation of several physical phenomena including Hofstadter's butterfly,\cite{Hunt2013,Dean2013,Ponomarenko2013} topological currents,\cite{Gorbachev2014} correlated insulator behaviour\cite{Cao2018a} and unconventional superconductivity.\cite{Cao2018} Critical to these observations is the formation of a commensurate state in which graphene is locally stretched in domains separated by sharp domain walls. Previous works have reported a commensurate-incommensurate transition at twist-angles $\theta$ (i.e. the angle formed between the lattice vectors of graphene and hBN) of the order of the lattice mismatch ($\sim1^{\circ}$).\cite{Woods2014} For $\theta<\delta$ graphene forms these domains of strong van der Waals interaction with hBN, whilst in the opposing case ($\theta>\delta$) local strain is not observed. Thermal annealing has been shown to induce an incommensurate-commensurate transition over micrometer scales providing the initial twist-angle is small ($\theta \leq 2^{\circ}$). For flakes which do not align, a 1D network of wrinkles emerges due to the difference in thermal expansion coefficients between hBN and graphene.\cite{Woods2016}

The quantum transport characteristics of twist-angle structures are commonly probed in transistor-like geometries. These generally consist of a graphene flake etched into a multi-terminal Hall-bar shape with metal contacts. However, the deposition of metal films onto graphene is known to induce structural defects, doping and strain\cite{Wang2011,Shioya2014,Shioya2015}. Metal contacts are possibly responsible for the failure of the devices upon thermal annealing, cooling at cryogenic temperatures or further processing such as encapsulation in ionic gates\cite{Ye2011,Mehew2017}. Ascertaining the role of the contacts on the properties of graphene/hBN superlattice structures is of pivotal importance and presently the focus of growing interest. For example, although phase- and growth-engineered 1D contacts have been explored in several atomically-thin materials\cite{Kappera2014,Cho625,Guimaraes2016}, the potential of 1D metallic edge-contacts for other encapsulated heterostructures has not yet been fully explored.

In this work we study the effect of strain induced by metal contacts in graphene/hBN superlattice devices. Semi-encapsulated graphene/hBN Hall-bars have been fabricated from a single flake with two different types of contact geometry: (1) two-dimensional (2D) top-contacts and (2) one-dimensional (1D) edge-contacts. Raman spectroscopy mapping was used to determine the strain and doping levels of the semi-encapsulated graphene. The absence of a top hBN layer allows to compare the two contact types in the same device. Top-contacts were found to induce strain levels up to $0.12\%$, localised between two opposing leads, whilst edge-contacts do not induce any measurable strain. Surprisingly electrical transport measurements in devices encompassing the two types of contact geometry showed different twist-angles depending on the contact type, despite being fabricated on the same flake. Post-processing thermal annealing is shown to reduce the overall doping across the device, whilst increasing the overall strain. At the same time, the twist-angle is shown to change with thermal annealing resulting in its convergence to the same value for both contacts type. Our results are supported by finite elements method (FEM) simulations which correctly predict the observed strain pattern induced in graphene/hBN by top contacts geometries. Our studies unveil the interplay between strain and contact geometry in semi-encapsulated graphene on hBN. This knowledge could be instrumental in experimentally accessing the rich physics arising from the introduction of a gauge potential in the effective Hamiltonian of two-dimensional materials, such as the realisation of a purely strain-based valley filter\cite{Fujita2010,Grujic2014} or a charge-funnelling device for energy applications\cite{DeSanctis2018}.

\section{Strain in G/hBN Hall-bar devices}
\begin{figure*}{}
	\centering
	\includegraphics[]{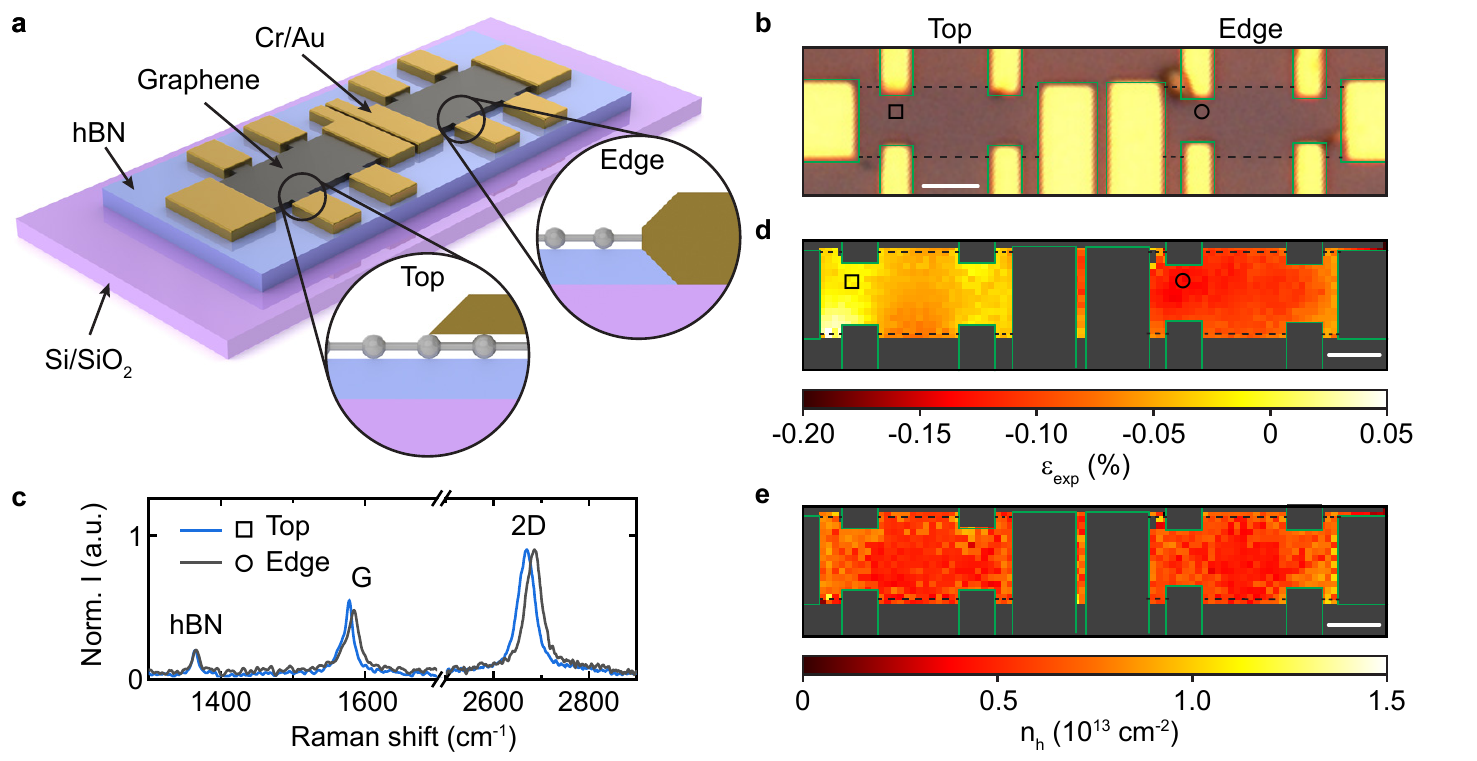}
	\caption{Strain patterns at metal contacts in graphene/hBN heterostructures. (a) Diagram of the investigated device with two types of metallic contacts, top and edge. (b) Optical micrograph of the device. (c) Raman spectra acquired in proximity of the top (square in panel (b)) and edge (circle in panel (b)) contacts showing a shift of the G and 2D peaks of graphene and no shift of the hBN peak at $1360\,\mathrm{cm^{-1}}$. (d) Experimentally determined strain, $\varepsilon_\mathrm{exp}$ and (e) hole doping, $n_\mathrm{h}$ maps across the device in panel (b). $\varepsilon<0$ indicates compressive strain. Scalebars are $2\,\mu\mathrm{m}$.}
	\label{fig:Figure1}
\end{figure*}

A single flake of monolayer graphene on hBN was used to fabricate Hall-bar devices (see methods) with two sets of metallic contacts, namely top- and edge-contacts, as schematically shown in \cref{fig:Figure1}a. The two sets of contacts have profound differences: edge electrodes make contact along a 1D chain of carbon atoms due to the etching step immediately before metal evaporation,\cite{Wang2013b} whilst top electrodes overlap with the graphene flake, effectively making a 2D (or planar) electrical contact.\cite{Allain2015} A micrograph of the device is shown in \cref{fig:Figure1}b. No top hBN layer was used to encapsulate the device as this would forbid the fabrication of 2D top-contacts and, if used only for 1D edge-contacts, will not allow a direct comparison of the strain and doping levels in the underlying graphene.

The Raman spectra acquired in the proximity of the two contacts (square and circle symbols in \cref{fig:Figure1}b) are shown in \cref{fig:Figure1}c. An immediate difference is visible: both the G and 2D modes of graphene ($\sim{1580}\,\mathrm{cm^{-1}}$ and $\sim{2670}\,\mathrm{cm^{-1}}$, respectively) are up-shifted for the edge- compared with top-contacts, whilst the $E_\mathrm{2g}$ phonon mode of hBN remains at $\sim{1350}\,\mathrm{cm{^{-1}}}$ in both cases. 

Raman maps (see Methods) of the whole device are used to determine the strain and doping configuration of the graphene Hall-bar. The technique used is based on the work of \citet{Lee2012a} and illustrated in detail in Figures S2 and S3, Supporting Information, here we briefly outline it. The frequency of the G ($\omega_\mathrm{G}$) and 2D ($\omega_\mathrm{2D}$) modes of graphene are extracted using a Lorentzian fit and plotted against each other. The data have a distribution which follows two principal axes crossing the pristine (unstrained and undoped) graphene coordinates ($\omega_\mathrm{G}^0=1581.6\,\mathrm{cm^{-1}}$, $\omega_\mathrm{2D}^0=2676.9\,\mathrm{cm^{-1}}$)\cite{Lee2012a}. These axes represent the iso-doping $(\Delta\omega_\mathrm{2D}/\Delta\omega_\mathrm{G})_\varepsilon = 2.2$ and the iso-strain $(\Delta\omega_\mathrm{2D}/\Delta\omega_\mathrm{G})_n = 0.70$ cases.\cite{Lee2012a} From this plot it is possible to correlate $\omega_\mathrm{G}$ and $\omega_\mathrm{2D}$ to the strain and (hole) doping levels by projecting each experimental point onto the principal axes and using the projected data to compute the strain and doping maps across the device.

Applying this method to the Hall-bar device in \cref{fig:Figure1}b we observe a clear distinction in the experimentally-determined level of strain ($\varepsilon_\mathrm{exp}$) between the two contact regions, \cref{fig:Figure1}d. For the edge-contacts compressive strain is uniformly distributed across the graphene channel with an average value of $\varepsilon_0\sim-0.12\,\%$, in agreement with literature.\cite{Neumann2015} Conversely, for the top-contacts region, compressive strain is relaxed between opposing electrodes, with larger compressive strain levels in the central area between electrodes. In contrast to strain, the doping ($n_\mathrm{h}$) is uniform across both top and edge contact geometries, \cref{fig:Figure1}e. A similar pattern is observed in a second device formed only by top-contacts, as shown in figure S4, Supporting Information. 

\begin{figure*}{}
	\centering
	\includegraphics[]{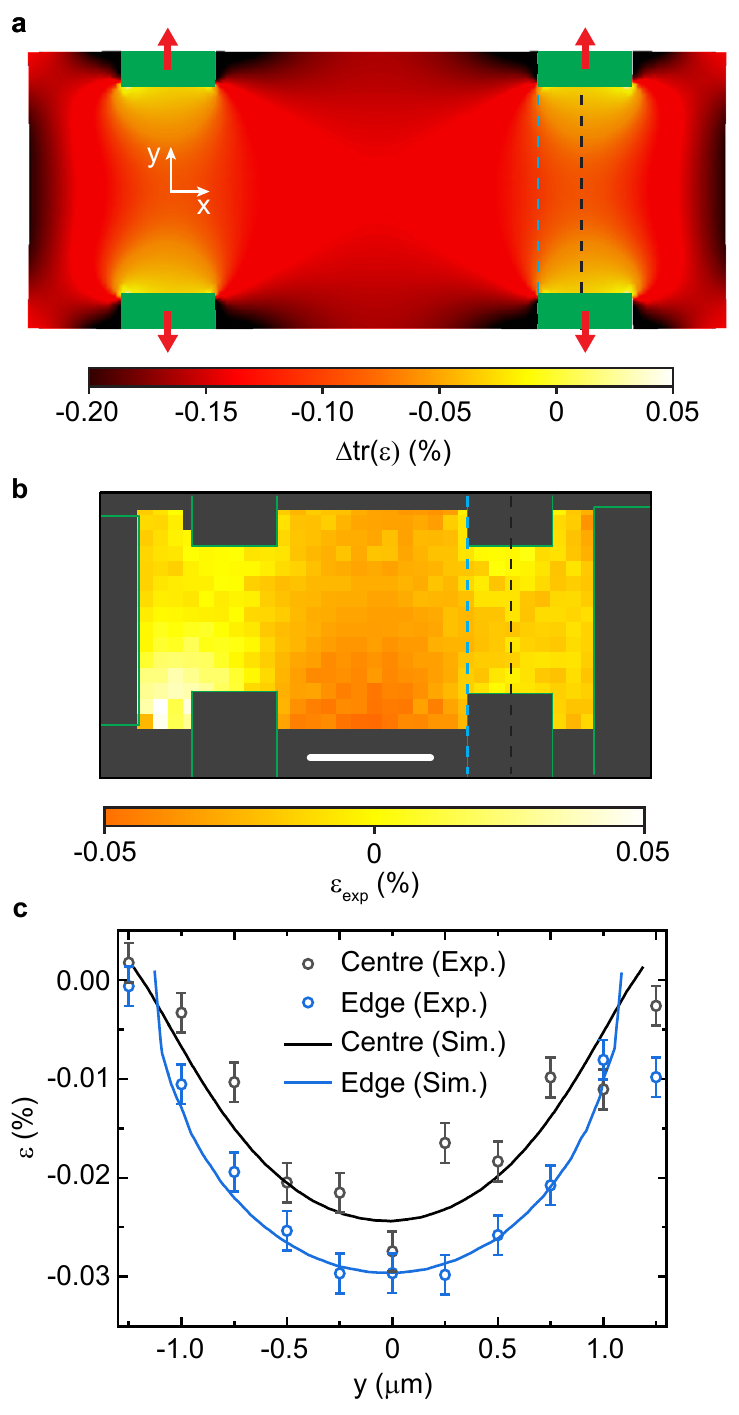}
	\caption{Finite elements modelling (FEM) simulation of strain patterns. (a) Simulated strain in a graphene membrane with top-contacts pulling in the direction of the red arrows. Colour-scale represents the trace of the simulated strain tensor $\varepsilon_\mathrm{xx}+\varepsilon_\mathrm{yy}$ plus the average initial strain $\varepsilon_0 = -0.12\,\%$ (see also main text),  $\varepsilon<0$ indicates compressive strain. (b) High resolution strain map of the graphene/hBN Hall bar with 2D top-contacts. (c) Experimental (dots) and simulated (solid lines) profiles along the blue and black dashed lines shown in panels (a) and (b). Scalebars are $2\,\mu\mathrm{m}$.}
	\label{fig:Figure2}
\end{figure*}

This experimental evidence suggests that top-contacts are ``pulling'' graphene, relaxing the existing compressive strain. To validate this idea, we perform a finite elements modelling (FEM) simulation and describe graphene as a two-dimensional, deformable membrane with an in-plane force applied normal to each contact region. \Cref{fig:Figure2}a shows the result of this analysis, where the difference between the simulated trace of the strain tensor and the initial compressive strain is plotted in a colour map ($\Delta\mathrm{tr}(\varepsilon)=(\varepsilon_\mathrm{xx}+\varepsilon_\mathrm{yy})-\left|\varepsilon_0\right|$). Relaxation of the initial compressive strain is observed at the contacts and, more interestingly, a bow-tie feature is observed between opposing contacts. Both these features are present in the experimentally measured strain map, \cref{fig:Figure2}b. \Cref{fig:Figure2}c compares the experiment and simulated strain values for line-cuts in \cref{fig:Figure2}a,b with reasonable agreement between the two.

The relaxation of compressive strain in graphene by the metal electrodes can be understood in the following way. Metal deposition elevates the device temperature which subsequently cools once evaporation is complete. Given the difference between the thermal expansion coefficients of Au (positive) and graphene (negative), upon cooling graphene expands whilst gold contracts. This contraction dominates as the thermal expansion coefficient of gold ($\sim{14\cdot10^{-6}}\,\mathrm{K^{-1}}$) is greater than that of graphene ($\sim{-7\cdot10^{-6}}\,\mathrm{K^{-1}}$).\cite{Bao2009,Yoon2011a} Therefore the contraction of the gold contacts relaxes the strained graphene with this relaxation extending into the channel area. While this is the case for top contacts, no strain relaxation is observed for edge contacts within the resolution of our experimental technique ($\sim0.005\,\%$), which can be explained by the order of magnitudes difference in the effective contact area between the two types of contacts. Edge-contacts in fully encapsulated devices are expected to induce similar strain level to the one observed in semi-encapsulated devices since the area of mechanical clamping to the graphene layer is the same.

\section{Effect of annealing on strain patterns}
\begin{figure*}{}
	\centering
	\includegraphics[]{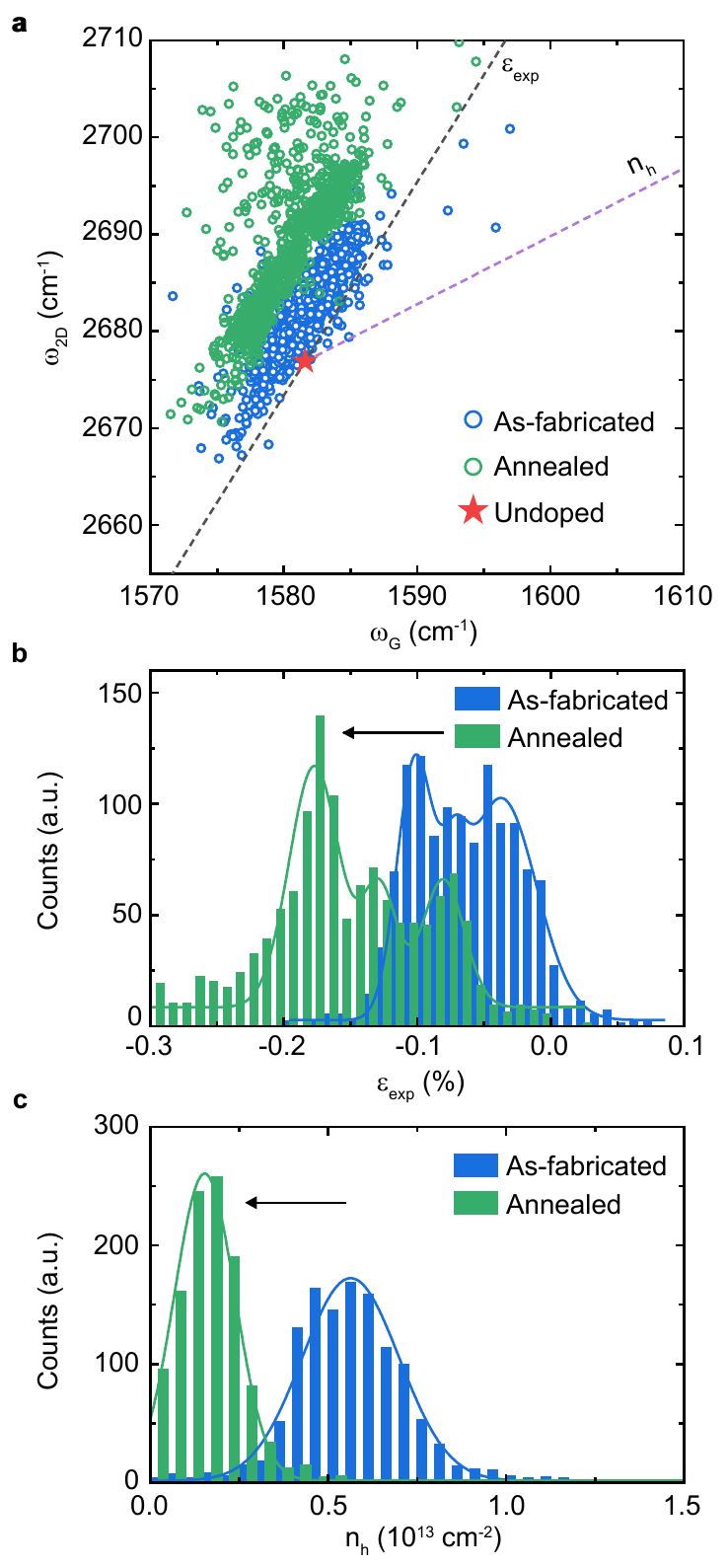}
	\caption{Effect of thermal annealing. (a) 2D \textit{vs} G peak frequency ($\omega_\mathrm{2D}$ and $\omega_\mathrm{G}$) for the pristine and annealed device. Dashed lines represent the iso-doping and iso-strain directions. (b) Statistical distributions of strain and (c) doping for the as-fabricated and annealed devices extracted from \ref{fig:Figure1}a,b and figure S5, Supporting Information. $\varepsilon<0$ indicates compressive strain. Solid arrow indicates an increase of compressive strain and reduction of overall doping after thermal annealing.}
	\label{fig:Figure3}
\end{figure*}

Thermal annealing is commonly used to enhance the electrical properties of graphene field-effect transistors by improving the metal-graphene interface and reducing contamination (e.g from polymer residues). However such procedure can lead to contact failure. In view of the observed strain pattern in our device, we study the effect of thermal annealing in this structure. \Cref{fig:Figure3}a shows the $\omega_\mathrm{G}$/$\omega_\mathrm{2D}$ distribution of the data from \cref{fig:Figure1}. In the as-fabricated device, the data are distributed along the iso-strain axis with a vertical shift away from the pristine case. This is thought to be due to Fermi velocity reduction, previously reported for graphene on hBN, which arises from van der Waals interlayer interaction.\cite{Hwang2012a,Ahn2013}. Upon thermal annealing for $2$ hours in forming gas (H$_2$/Ar, $10\%/90\%$) at $200\,\mathrm{^\circ C}$ the data-set shifts vertically upwards, suggesting a greater Fermi velocity reduction from increasing interlayer interaction (a strain/doping map of the device after thermal annealing is shown in figure S5, Supporting Information). Thermal treatment has a pronounced effect on the strain distribution, as shown in \cref{fig:Figure3}b. In the map shown in \cref{fig:Figure1}d it is possible to identify three distinct regions with different average strain levels, two coming from the top-contact region (strain emerging from opposing top contacts and the channel between adjacent pairs) and one from the edge-contact region (uniform strain). This suggests that the statistical distribution of the data (\cref{fig:Figure3}b) should be modelled with three Gaussian distributions. We confirm this observation by analysing the data extrapolated from each region, as detailed in figure S6, Supporting Information, where two distinct distributions are observed to arise from the top-contacts and one from the edge-contacts. Upon annealing, strain increases in all areas as evidenced by the values reported in \cref{tab:strain}. Strain in the edge-contacted region increases from $-0.102\,\%$ to $-0.177\,\%$ ($\Delta\varepsilon\sim0.07\,\%$) indicating that graphene has become more compressed. Similar compression is observed for top-contacts where close to (away from) the electrode an increase of $\Delta\varepsilon\sim0.05\,\%$ ($\Delta\varepsilon\sim0.06\,\%$) is extracted. 

Previous reports have shown that graphene on hBN can undergo a rotation upon thermal annealing,\cite{Woods2016} increasing the crystallographic alignment, which occurs as the system tries to minimise the interlayer van der Waals energy. This suggests that there is a competition between flake rotation and mechanical clamping from the metal electrodes. With increased clamping from top contacts a smaller change in strain occurs in these regions. This observation highlights the potential negative role played by this phenomenon on the failure of contacts in graphene/hBN devices.
 
\begin{table}[t]
	\begin{center}
		\begin{tabular}{l c c c c}
			\hline
			& \multicolumn{2}{c}{$\varepsilon_\mathrm{exp}$ (\%)} & \multicolumn{2}{c}{$n_h\,(10^{12}\,\mathrm{cm^{-2}})$} \\ \cline{2-5}
			Contact Type & As-fabricated & Annealed & As-fabricated & Annealed \\ \hline
			Edge & $-0.102\pm0.002$ & $-0.177\pm0.002$ & $5.70\pm0.03$ & $1.44\pm0.05$ \\ 
			Top (contact) & $-0.037\pm0.001$ & $-0.090\pm0.003$ & $=^\mathrm{a}$ & $=$\\
			Top (channel) & $-0.057\pm0.005$ & $-0.128\pm0.003$ & $=$ & $=$ \\
			\hline
		\end{tabular}
	\end{center}
	\caption[Strain Values]{Comparison of strain and doping values before and after annealing extrapolated from the Gaussian fits in \cref{fig:Figure3}b,c. Top (contact) refers to the region between two opposing metal contacts. Top (channel) refers to the graphene region between two adjacent metal contacts (see \cref{fig:Figure1}).}
	$^\mathrm{a}$ Same value as previous.
	\label{tab:strain}
\end{table} 

\Cref{fig:Figure3}c shows a histogram of the doping levels used as an indication of contamination. In contrast to the strain statistics, a single Gaussian can be fitted to the data indicating uniform doping across both contact regions with $n\sim{5.6\cdot10^{12}}\,\mathrm{cm^{-2}}$. As expected, following annealing this reduces to $n\sim{1.5\cdot10^{12}}\,\mathrm{cm^{-2}}$, validating the usefulness of this common processing step in enhancing the electrical properties of graphene devices.

\begin{figure*}{}
	\centering
	\includegraphics[]{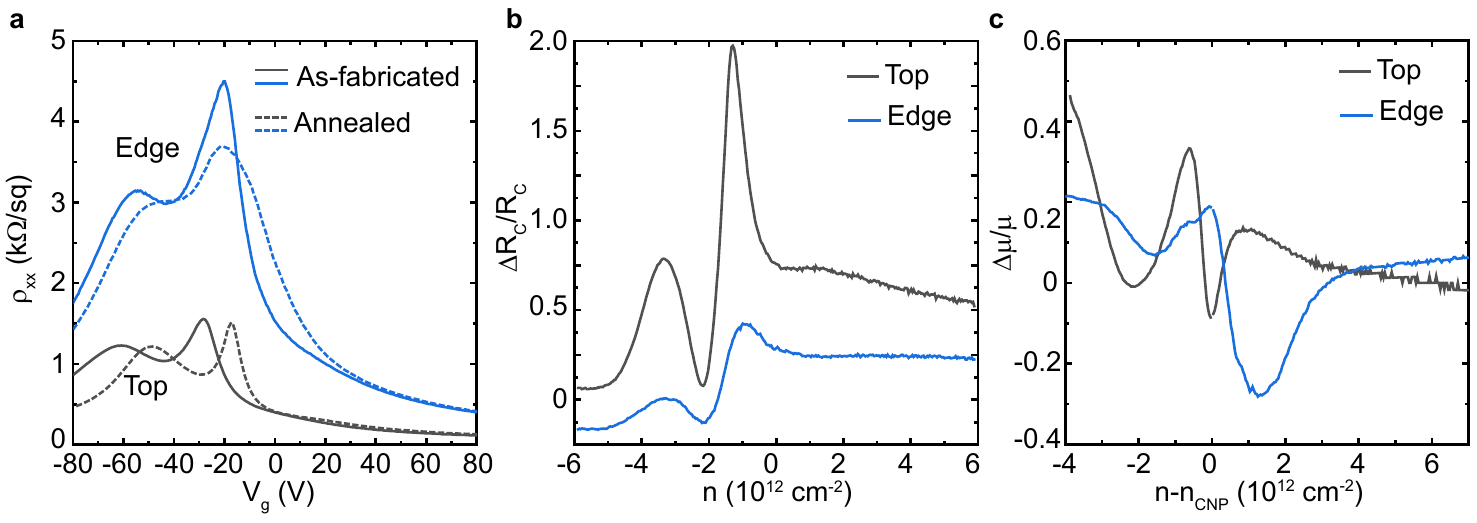}
	\caption{Effect of annealing on the electrical properties of top- and edge-contacted devices. (a) Longitudinal resistivity $\rho_\mathrm{xx}$ \textit{vs} gate voltage $V_\mathrm{g}$ for the two contact types. Solid lines correspond to the as-fabricated device and dashed lines correspond to the annealed device. (b) Normalised contact resistance difference ($\Delta R_\mathrm{c}/R_\mathrm{c}$) between as-fabricated and annealed devices for the two contact types as a function of induced carrier density $n$. (c) Normalised mobility difference ($\Delta \mu/\mu$) between as-fabricated and annealed devices for the two contact types as a function of carrier density $n-n_\mathrm{CNP}$, where $n_\mathrm{CNP}$ is the carrier density to reach the charge neutrality point.}
	\label{fig:Figure4}
\end{figure*}

Electrical measurements help to shed light on the role of thermal annealing on strain for the different contact configurations. \Cref{fig:Figure4}a shows the resistivity ($\rho_\mathrm{xx}$) as a function of gate voltage ($V_\mathrm{g}$) measured in the two contacts regions before and after thermal annealing. For both contact types, a down-shift of the charge neutrality point (CNP) after annealing is observed, with corresponding doping values comparable to those extrapolated from Raman spectroscopy. The change in contact resistance between the as-fabricated and annealed devices $\Delta R_\mathrm{c}/R_\mathrm{c}$, where $R_\mathrm{c}=\left(R_\mathrm{2T}-R_\mathrm{4T}\right)/2$, and the corresponding change in mobility $\Delta\mu/\mu$ are shown in \cref{fig:Figure4}b and \cref{fig:Figure4}c, respectively. An increase by as much as $200\%$ of the contact resistance following annealing for top-contacts was observed, whereas a change of $\sim30\%$ is observed for edge-contacts. This is in line with the observation that annealing top-contacts in the presence of strain can lead to their electrical failure. In top-contacts the presence of strain, due to the difference in thermal expansion between the metal and graphene, is an indication of good mechanical clamping. These contacts always result in good electrical connections. On the other hand, annealing of top-contacts with poor mechanical clamping which do not show measurable levels of strain in graphene result in open electrical contacts, see Supporting Information figure S7. This is in stark contrast to the observations reported on the edge-contacts which always result in good electrical connections in the absence of any measurable strain. Thermal annealing slightly improves the charge carrier mobility by the same amount ($\sim20\%$) in both contact geometries, see \cref{fig:Figure4}c. This indicates that the charge carrier mobility is limited by Coulomb impurities and not by strain-induced modifications to the deformation potential acoustic phonon scattering.

\section{Twist-angle in strained G/hBN superlattices}
Two peaks in resistivity are observed in \cref{fig:Figure4}a. The CNP appears at $V_g\sim-20\,\mathrm{V}$ ($\rho_\mathrm{xx}=4.5\,\mathrm{k\Omega/sq}$) whilst a second, satellite peak, appears at $V_{g}\sim{-60}\,\mathrm{V}$. This second peak arises due to the emergence of additional Dirac points in the band structure of graphene on hBN, as previously observed in low-temperature transport experiments\cite{Yankowitz2012,Yang2013,Hunt2013,Dean2013,Ponomarenko2013} and more recently at room temperature\cite{Tang2015,Ribeiro-Palau2018} (see also figure S8, Supporting Information, for measurements on a second device).

\begin{figure*}{}
	\centering
	\includegraphics[]{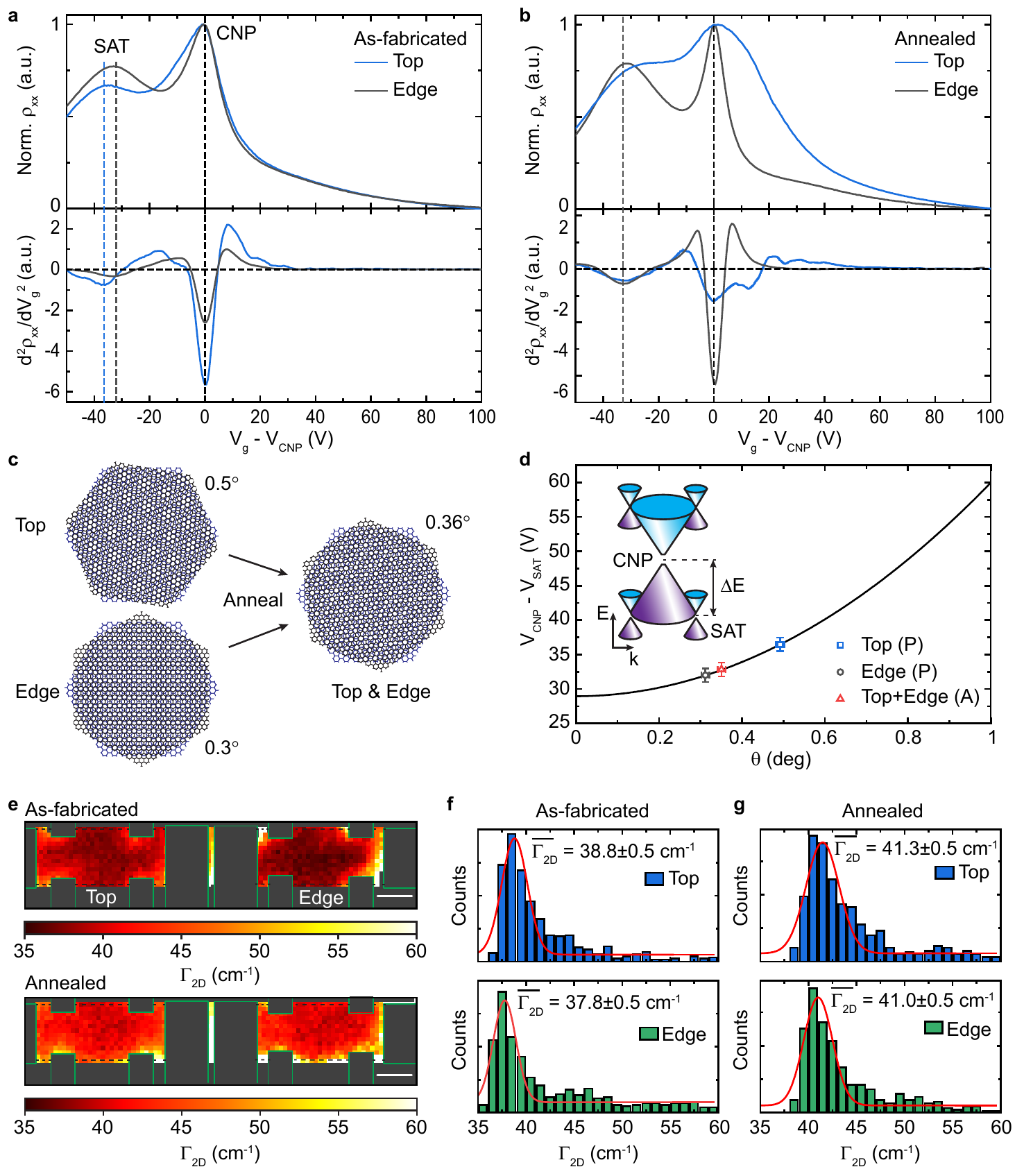}
	\caption{Graphene/hBN superlattices. (a) Normalised longitudinal resistivity $\rho_\mathrm{xx}$ (top) and its second derivative with respect to gate voltage (bottom) as a function of $V_\mathrm{g} - V_\mathrm{CNP}$ for the as-fabricated device. (b) Same quantities measured in (a) after annealing. (c) Schematic diagrams showing the Moir\'e superlattices for top- and edge-contacts before and after annealing. (d) Measured (dots) positions of the satellite ($V_\mathrm{SAT}$) and charge-neutrality points ($V_\mathrm{CNP}$) and corresponding angles of rotation from \cref{eq:vgsat} (solid line). Inset: illustration of satellite mini-bands arising from the Moir\'e superlattice. (e) Maps of the full-width-at-half-maximum (FWHM) of the 2D band of graphene ($\Gamma_\mathrm{2D}$) before (top) and after (bottom) thermal annealing of the device shown in \cref{fig:Figure1}c. (f) and (g) Statistical distributions of $\Gamma_\mathrm{2D}$ from the map in panels (a) and (b) for the two different contacts regions. Red solid lines correspond to Gaussian fits centred at $\overline{\Gamma_\mathrm{2D}}$. Scale-bars are $2\,\mathrm{\mu m}$.}
	\label{fig:Figure5}
\end{figure*}

The interaction between hBN and graphene, which modifies the band structure of the latter, is tuned by the crystallographic angle between the two materials. This is reflected in a change of the separation of these two peaks in transport measurement which can be correlated to this angle.\cite{Ribeiro-Palau2018} This separation is examined more closely in \cref{fig:Figure5}a,b. In the as-fabricated device the satellite peak ($V_\mathrm{SAT}$) is at more negative gate voltages for edge regions. This can be clearly seen by taking the second derivative of the resistivity ($d^2\rho_\mathrm{xx}/dV_\mathrm{g}^2$) indicating that different twist angles exist in the top and edge contact regions, a difference which disappears after annealing, \cref{fig:Figure5}b. Increasing the twist angle from $\theta={0^{\circ}}$ reduces the Moir\'e wavelength ($\lambda$) and manifests as a shift of the satellite Dirac points away from the main Dirac point (see \cref{fig:Figure5}d inset). The Moir\'e wavelength can be expressed as:\cite{Yankowitz2012,Ribeiro-Palau2018}

\begin{equation}
	\lambda = \frac{\left(1+\delta\right)a}{\sqrt{2\left(1+\delta\right)\left[1-\cos\theta\right]+\delta^2}},
	\label{eq:moire}
\end{equation}
where $\delta\sim0.017$ is the lattice mismatch between graphene and hBN and $a=0.246\,\mathrm{nm}$ is the lattice constant of graphene. Due to the spin and valley degeneracies in graphene, full-filling occurs at a density of four electrons per superlattice cell ($n=4n_0$), with the unit cell area $1/n_0=\sqrt{3}\lambda^2/2$.\cite{Hunt2013} Since the carrier density is $n=C_\mathrm{g}(V_\mathrm{g}-V_\mathrm{CNP})/e$ where $C_\mathrm{g}$ is the geometric gate capacitance, $V_\mathrm{CNP}$ is the position of the charge-neutrality point and $e$ is the electron charge, we find:

\begin{equation}
	(V_\mathrm{g}-V_\mathrm{CNP}) = \frac{8e}{\sqrt{3}\lambda^2 C_\mathrm{g}},
	\label{eq:vgsat}
\end{equation}

Using \cref{eq:moire} and \cref{eq:vgsat} it is therefore possible to extrapolate the twist-angle between graphene and hBN from the data in \cref{fig:Figure5}a,b. \Cref{fig:Figure5}c is a schematic illustration of the Moir\'e superlattice structure formed by rotating the graphene with respect to the hBN. Initially two twist angles are present, $\theta_\mathrm{Top}=(0.312\pm0.005)^\circ$ and $\theta_\mathrm{Edge}=(0.492\pm0.005)^\circ$. Upon thermal annealing, the heterostructure self-reorientates and one angle is measured for both regions ($\theta=(0.351\pm0.005)^\circ$). In \cref{fig:Figure5}d the experimental data has been plotted with the functional dependence of $V_\mathrm{CNP}-V_\mathrm{sat}$ against twist angle. Here, the free parameter is $C_\mathrm{g}=1.18\cdot10^{-4}\,\mathrm{Fm^{-2}}$ which has been estimated by considering the two dielectrics (hBN and SiO$_2$) stacked in series.

The Moir\'e wavelength is related to the full-width at half-maximum (FWHM) of the 2D peak ($\Gamma_\mathrm{2D}$) by: $\Gamma_\mathrm{2D}=2.7\lambda+0.77$, where the numerical constant ($0.77$) is dependent on the device structure and electrical properties (e.g. mobility)\cite{Ribeiro-Palau2018}. Given that this constant is unknown in our devices we cannot directly compare the absolute value of $\lambda$ extrapolated from Raman and transport measurements. We are however able to confirm the convergence of the twist-angle after thermal annealing. \Cref{fig:Figure5}e shows the maps of $\Gamma_\mathrm{2D}$ obtained from the device in \cref{fig:Figure1}b before and after thermal annealing. The statistical distribution of $\Gamma_\mathrm{2D}$ for the two contact regions, before and after annealing, is shown in \cref{fig:Figure5}f,g. Before thermal annealing we obtain a difference in $\overline{\Gamma_\mathrm{2D}}$ (centre of the Gaussian distribution) of $\Delta\overline{\Gamma_\mathrm{2D}}=1.0\pm0.5\,\mathrm{cm^{-1}}$. After annealing this value reduces to $\Delta\overline{\Gamma_\mathrm{2D}}\simeq0\,\mathrm{cm^{-1}}$, confirming the convergence of the twist-angle observed in electronic transport measurements. Furthermore, the absolute values of $\overline{\Gamma_\mathrm{2D}}$ after thermal annealing ($\sim41\,\mathrm{cm^{-1}}$) are higher than the corresponding values before annealing ($\sim37.5\,\mathrm{cm^{-1}}$), suggesting that the twist-angles converge to a higher value, in apparent contrast with the data shown in \cref{fig:Figure5}d. Indeed, this is not the case as both the FWHM of the G peak ($\Gamma_\mathrm{G}$) and the value of $\Gamma_\mathrm{2D}$ depend on strain\cite{Neumann2015}, with a slope $(\Delta\Gamma_\mathrm{2D}/\Delta\Gamma_\mathrm{G})_\varepsilon=2.2$. Therefore, the increase in $\overline{\Gamma_\mathrm{2D}}$ after annealing confirms the data shown in \cref{fig:Figure3}a,b and supports the observation that thermal annealing increases the overall compressive strain.

\section{Summary and discussion}
To summarise, we have analysed the strain induced in single-layer graphene deposited on hBN by 2D top-contacts and 1D edge-contacts. Using Raman spectroscopy mapping, we have shown that top-contacts induce strain in the graphene flake, pulling in opposite directions. On the contrary, edge-contacts do not induce such strain. Our observation is supported by FEM simulations. We associate this strain to the shrinking of Au and graphene which, due to their different thermal expansion coefficients, lead to a net pull by the metal contacts after thermal evaporation of the metal. Thermal annealing on such devices reduces the overall doping, as expected from the removal of contaminants, but it also increases the overall compressive strain on the graphene. Such increase can lead to contacts failure given the observed strain pattern on the flake. Furthermore, on aligned samples, where the graphene and hBN lattice vectors are rotated by a small angle ($\theta<1\,^\circ$), we have shown that the angle $\theta$ is different for the two types of contacts, although it converges to the same angle after thermal treatment. The convergence of the twist-angle confirms the self-locking mechanism observed in aligned graphene/hBN heterostructures\cite{Woods2016}. The interplay between rotation and contacts-induced strain can also lead to contacts failure. Full encapsulation in hBN with edge-contacts may affect the way the angle $\theta$ changes upon annealing, however the negligible levels of strain observed in the edge-contacts suggest that this would not play a major role in this kind of devices.

Our observations elucidate the role of metal contacts on inducing strain in 2D materials-based electronic heterostructures. These findings can have profound implications on the development of future electronic applications based on layered two-dimensional materials in sensing and quantum computing. Metal contacts can be used to engineer complex strain patterns which could provide a system in which new physical phenomena could be investigated. The introduction of a gauge potential in the effective Hamiltonian, for example, can be used to realise a purely strain-based valley filter\cite{Fujita2010,Grujic2014}, whilst non-uniform bandgap modulation in semiconducting transition-metal dichalcogenides can be used in charge-funnelling devices for energy harvesting\cite{DeSanctis2018}.

\section*{Methods}

\subsection*{Samples fabrication}
Hexagonal Boron Nitride supplied by \textit{Manchester Nanomaterials} was mechanically exfoliated onto SiO$_2$/Si$^+$ substrates that had been previously treated with a high power ($30\,\mathrm{W}$) O$_2$ plasma. This processing step increases both the yield and lateral size of exfoliated flakes. Graphene was exfoliated onto a polymer bilayer (PMMA/PVA) and placed on the hBN by dry transfer.\cite{Pizzocchero2016} The lithography steps performed to create both top- and edge-contacts on the same flake are described in figure S1 and associated Supporting text. Electron beam lithography (\textit{Nano Beam} NB5) was performed using $500\,\mathrm{nm}$-thick PMMA (\textit{MicroChem} 950K A6), developed using a 3:1 solution of isopropanol (IPA) and 4-Methyl-2-pentanone (MIBK). Contacts were deposited using Cr/Au ($15/60\,\mathrm{nm}$) thermal evaporation at pressure $<5\cdot10^{-7}\,\mathrm{Torr}$ and lift off in hot Acetone ($2\,\mathrm{hrs}$), then left overnight in Acetone. For edge-contacts, prior to metal deposition the exposed contact areas of the graphene/hBN stack were etched using a CHF$_3$/O$_2$ reactive-ion plasma at pressure $P = 30\,\mathrm{mT}$ with $40\,\mathrm{sccm}$ of CHF$_3$ plus $4\,\mathrm{sccm}$ of O$_2$, power $25\,\mathrm{W}$ for $30\,\mathrm{sec}$.

\subsection*{Electrical transport measurements}
AC lock-in measurement techniques (\textit{Ametek} Signal Recovery 7270) were employed to accurately probe changes in resistivity with small excitation voltages ($V_{ac}\sim{1}\,\mathrm{mV}$) minimising Joule heating in the device in a  custom-built measurement chamber\cite{DeSanctis2018_RSI}. The excitation voltage was modulated at a frequency of $72.148\,\mathrm{Hz}$. Two- ($V_\mathrm{2T}$) and four-terminal ($V_\mathrm{4T}$) voltages allowed the simultaneous measurement of channel resistivity, $\rho_\mathrm{xx}$, field-effect mobility, $\mu$, and contact resistance, $\left(R_\mathrm{2T}-R_\mathrm{4T}\right)/2$. The graphene channel was capacitively coupled to the Si$^{+}$ backgate through a $280\,\mathrm{nm}$-thick SiO$_2$ layer, allowing the modulation of carrier density $n$ by applying a DC voltage. All measurements were performed in vacuum ($P<10^{-6}\,\mathrm{mBar}$) at room temperature.

\subsection*{Raman spectroscopy}
Raman spectra were acquired using a custom-built set-up with a $514\,\mathrm{nm}$ excitation CW solid-state laser as source, focussed thorough a $\times50$ lens (NA$=0.9$, \textit{Olympus} MPLFLN).\cite{DeSanctis2017} Back-scattered light was collected by the same lens and, after filtration of the excitation line, dispersed by a $1800\,\mathrm{g/mm}$ grating mounted in a \textit{Princeton Instruments} Acton SP2500 spectrometer and the spectra measured by a \textit{Princeton Instruments} PIXIS400 CCD camera. The laser spot-size is $484\,\mathrm{nm}$, compatible with the lateral step-size of the motorised stage ($500\,\mathrm{nm}$)\cite{DeSanctis2017}. All measurements were performed in vacuum ($P<10^{-6}\,\mathrm{mBar}$) at room temperature.

We have tested our system to perform the analysis outlined in the text in order to confirm the ability to resolve strain and doping levels, as detailed in figure S2. As shown in figure S3, for our test we employed a graphene flake deposited half on hBN and half on SiO$_2$. Our analysis correctly showed a difference in both strain and doping in these two regions and highlighted an average compressive strain for graphene on hBN of $\varepsilon_0\sim-0.12\,\%$, in agreement with literature.\cite{Neumann2015} 

\subsection*{FEM simulations}
FEM simulations were performed using Elmer open source multiphysical simulation software freely available from \url{http://www.elmerfem.org}. The built-in linear elasticity model solver was used. Graphene was simulated as a 2D membrane with a mesh containing $241$ nodes and $8154$ surface elements. The physical size of the mesh was proportional to the actual device, shown in \cref{fig:Figure1}c, with a linear scaling factor of $133\cdot10^{3}$. The following elasticity parameters where used\cite{Politano2015}: in-plane Young's modulus $E=340\,\mathrm{N/m}$ and Poisson's ratio $\nu=0.165$. An in-plane force equivalent to $15\,\mathrm{\mu N}$ was applied at opposing contacts in opposite directions. This is in good agreement with the stress-strain curve of graphene\cite{Fthenakis2015}.

\section*{Acknowledgment}
We thank G. F. Jones for designing and building the vacuum chamber used in the Raman Spectroscopy measurements. M.F.C. and S.R. acknowledge financial support from: Engineering and Physical Sciences Research Council (EPSRC) of the United Kingdom, projects EP/M002438/1, EP/M001024/1, EPK017160/1, EP/K031538/1, EP/J000396/1; the Royal Society, grant title "Room temperature quantum technologies" and "Wearable graphene photovolotaic"; Newton fund, Uk-Brazil exchange grant title "Chronographene" and the Leverhulme Trust, research grants "Quantum drums" and "Quantum revolution". J.D.M. acknowledges financial support from the Engineering and Physical Sciences Research Council (EPSRC) of the United Kingdom, via the EPSRC Centre for Doctoral Training in Metamaterials, Grant No. EP/L015331/1. S.A. acknowledges financial support from the Higher Committee for Education Development in Iraq (HCED).

\section*{Author Contributions}
A.D.S. and J.D.M. contributed equally to this work. A.D.S. and S.R. conceived the idea. A.D.S. performed the measurements, FEM simulations, interpreted the data and wrote the manuscript. J.D.M. fabricated the devices, interpreted the data and wrote the manuscript. S.A. fabricated the devices for initial studies and interpreted the data. F.W. developed the 2D material transfer system and demonstrated the transfer methods used for the production the graphene-hBN heterostructures. M.F.C. and S.R. supervised the project and wrote the paper. 

\section*{Competing interests}
The authors declare no competing financial interests.

\begin{suppinfo}
Supporting Information is submitted in conjunction with this manuscript. Correspondence should be addressed to S.R. or A.D.S.
\end{suppinfo}

\bibliography{Strain-Contacts_ADeS_PAPER-biblio}

\end{document}